\documentclass[preprint,12pt]{elsarticle}%
\usepackage{amssymb}
\usepackage{amsfonts}
\usepackage{amsmath}
\usepackage{graphicx}%
\providecommand{\U}[1]{\protect\rule{.1in}{.1in}}
\graphicspath{{G:/ANTONIO/dottorato_(HD)/Dimple/Testo/figure/}}
\journal{journal}

\begin{document}
\bigskip%
\begin{frontmatter}%


%

\title
{A simplified theory of "stickiness" due to electroadhesion between rough surfaces}%

%

\author{M. Ciavarella}%
%

\address{Politecnico di BARI. DMMM dept. V Japigia 182, 70126 Bari. }
\address{email: mciava@poliba.it }%
%

\begin{abstract}%

Building on theories of Persson, we derive a simpler theory for
electroadhesion between rough surfaces using BAM (Bearing Area Model) of
Ciavarella, or previous ideas by Persson and Tosatti.\textit{ }Rather
surprisingly, in terms of stickiness, we obtain very simple and similar
results for pure power law power spectrum density (PSD), confirming stickiness
to be mainly dependent on macroscopic quantities. We define a new
dimensionless parameter for electroadhesive stickiness.%

\end{abstract}%
%

%

\end{frontmatter}%



\section{Introduction}

Applied electric voltage between two rough solids leads to accumulation of
charges of opposite sign on the surfaces, which obviously results in an
electrostatic attraction, which adds to the external repulsive loads (omitting
for simplicity the presence of van der Vaals independent forces). Also, this
increases the area of contact between the solids, and therefore the friction
force if solids are sliding, with particular reference to touch screen
applications (Vardar et al., 2017). Persson (2018) has extended his well known
theory of contact mechanics to the case of electroadhesion, and developed a
general mean-field theory in two limiting cases: (i) when an electric
insulating film separate two conducting bodies and (ii) two electric
conducting solids, which results in contact resistance and in voltage drop $V$
over a narrow region at the interface, using the theorem by Barber (2013),
that contact resistance is proportional to the mechanical contact stiffness.
Simplified results were obtained also by Popov \& Hess (2018).

As suggested in a very interesting recent paper by Dalvi \textit{et al.}
(2019), surface topography can easily have more than seven orders of magnitude
of almost power law spectrum, including down to the \AA ngstr\"{o}m-scale (see
Fig.S2 in Dalvi \textit{et al.} (2019) which gives the 2D isotropic PSD). The
very interesting recent paper by Dalvi \textit{et al.} (2019) reports
extensive adhesion measurements and corroborate ideas originally suggested by
Persson \&\ Tosatti (2001) and reworked in terms of stickiness criteria by
Ciavarella (2019). We shall extend this to electroadhesion, including the
stickiness criteria, and we shall find surprisingly universal results, despite
the very different origin of the various proposals we compare.

\section{Electro adhesion}

Consider an elastic solid with surface roughness above a rigid solid with a
flat surface. Both solids are conducting materials with insulating surface
layers of thickness $d_{1}$ and $d_{2}$ and dielectric constants
$\varepsilon_{1}$ and $\varepsilon_{2}$. An electric voltage difference $V$
occurs between the two conducting solids. The interfacial separation
$u=u\left(  \boldsymbol{x}\right)  $ depends on the lateral coordinate
$\boldsymbol{x=}\left(  x,y\right)  $.

\begin{center}
$%
\begin{array}
[c]{c}%
{\includegraphics[
height=1.5575in,
width=4.7165in
]%
{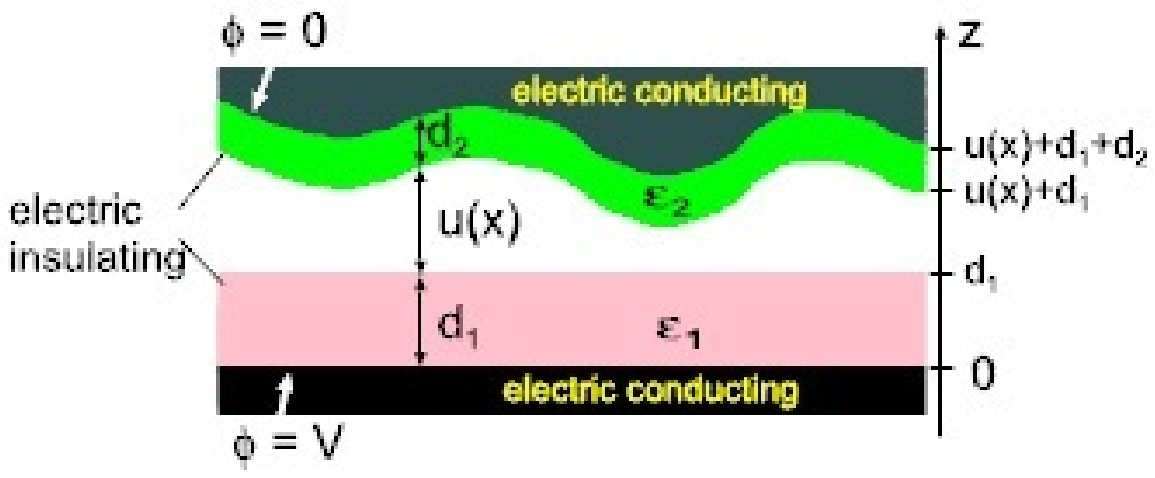}%
}
\\
\end{array}
$

Fig.2- Basic geometry of the problem, an elastic solid with surface roughness
above a rigid solid with a flat surface. Both solids are conducting materials
with insulating surface layers of thickness $d_{1}$ and $d_{2}$ and dielectric
constants $\varepsilon_{1}$ and $\varepsilon_{2}$. An electric voltage
difference $V$ occurs between the two conducting solids. The interfacial
separation $u=u\left(  \boldsymbol{x}\right)  $ depends on the lateral
coordinate $\boldsymbol{x=}\left(  x,y\right)  $.
\end{center}

With roughness, Persson finds using his contact mechanics theory an equation
which gives the real contact area and the nominal pressure as a function of
voltage. From his Fig.1 we see that the area becomes highly non linear with
voltage and also pressure.

Additionally, this "real contact" area remains highly ill-defined, depends
critically on the truncation of the PSD spectrum of roughness.

From eqt.just above (2) in Persson (2018), we have for the case of insulating
solids
\begin{equation}
\sigma_{zz}\left(  \boldsymbol{x}\right)  =\frac{\varepsilon_{0}}{2}\left(
\frac{V}{u\left(  \boldsymbol{x}\right)  +h_{0}}\right)  ^{2} \label{sigma}%
\end{equation}
where
\begin{equation}
h_{0}=\frac{d_{1}}{\varepsilon_{1}}+\frac{d_{2}}{\varepsilon_{2}}%
\end{equation}
is a fixed quantity. For the case of insulating solids, hence, we have an
adhesive term $p_{ad}=\left\langle \sigma_{zz}\left(  \boldsymbol{x}\right)
\right\rangle $ and being $p_{rep}$ the repulsive pressure, the final external
pressure results in (notice that it is negative when tensile),
\begin{equation}
\sigma=p_{rep}-p_{ad}=p_{rep}-\frac{\varepsilon_{0}}{2}V^{2}\int_{0}^{\infty
}\left(  \frac{1}{u\left(  \boldsymbol{x}\right)  +h_{0}}\right)  ^{2}P\left(
p,u\right)  du
\end{equation}
where $P\left(  p,u\right)  $ is the distribution of interfacial separations
computed by Persson theory in Almquist et al. (2011).

In what follows, we shall take a typical power law PSD
\begin{equation}
C\left(  q\right)  =Zq^{-2\left(  1+H\right)  }%
\end{equation}
for $q>q_{0}=\frac{2\pi}{\lambda_{L}}$, where $\lambda_{L}$ is the longest
wavelenght in the roughness spectrum, $H$ is the Hurst exponent (equal to
$3-D$ where $D$ is the fractal dimension of the surface), and specifically,
$Z=\frac{H}{2\pi}\left(  \frac{h_{0}}{q_{0}}\right)  ^{2}\left(  \frac
{1}{q_{0}}\right)  ^{-2\left(  H+1\right)  }$ where $h_{0}^{2}=2h_{rms}^{2}$.
Here, $h_{rms}$ is the rms amplitude of roughness.

\section{Idea 1 -- BAM method}

\bigskip

Instead of integrating the full distribution, we can use the idea of BAM
(Ciavarella, 2018), which gives a very simple estimate of adhesion of hard
solids with rough surfaces based on a bearing area model. Suppose we neglect
van der Waals adhesion, we just replace the true expression of the stress as a
function of separation (\ref{sigma}) with a Maugis-Dugdale equivalent for
which the tensile stress is defined as a function of gap $u$ as%

\begin{equation}%
\begin{tabular}
[c]{ll}%
$\sigma_{ad}\left(  u\right)  =\sigma_{0},$ & $u\leq h_{0}$\\
$\sigma_{ad}\left(  u\right)  =0,$ & $u>h_{0}$%
\end{tabular}
\ \ \ \ \ \qquad\qquad\qquad\qquad\label{maugis}%
\end{equation}

Upon integration for nominally flat surfaces,%
\begin{equation}
\Delta\gamma=\frac{\varepsilon_{0}}{2}V^{2}\int_{0}^{\infty}\left(  \frac
{1}{u\left(  \boldsymbol{x}\right)  +h_{0}}\right)  ^{2}du=\frac
{\varepsilon_{0}}{2h_{0}}V^{2}%
\end{equation}
and for this to be equal to the BAM corresponding integral $\Delta\gamma
=\int_{0}^{h_{0}}\sigma_{0}du=\sigma_{0}h_{0}$, we need to have
\begin{equation}
\sigma_{0}=\frac{\varepsilon_{0}}{2h_{0}^{2}}V^{2}%
\end{equation}
after which we can use BAM (Ciavarella, 2018).

For a Gaussian nominally flat surface, this results in
\begin{equation}
\frac{A_{ad}}{A_{0}}=\frac{1}{2}\left[  Erfc\left(  \frac{\overline
{u}-\epsilon}{\sqrt{2}h_{rms}}\right)  -Erfc\left(  \frac{\overline{u}}%
{\sqrt{2}h_{rms}}\right)  \right]
\end{equation}
where $\overline{u}$ is the mean separation of the surfaces, $h_{rms}$ is rms
amplitude of roughness. The total force is obtained by superposition of the
repulsive pressure at indentation $\Delta$ which is easily obtained with
Persson's theory (Persson, 2007) which, for the simplest power law PSD, and
$D\simeq2.2$ gives%
\begin{equation}
\frac{p_{\mathrm{{rep}}}\left(  \overline{u}\right)  }%
{E^{\raisebox{0.7mm}{$ *$}}}\simeq q_{0}h_{\mathrm{{rms}}}\exp\left(
\frac{-\overline{u}}{\gamma h_{\mathrm{{rms}}}}\right)  \label{Persson1}%
\end{equation}
where $\gamma\simeq0.5$ is a corrective factor. Therefore, summing up
repulsive\ (\ref{Persson1}) and attractive ($\sigma_{0}A_{\mathrm{{ad}}%
}(\overline{u})$) contributions, BAM gives
\begin{equation}
\frac{\sigma\left(  \overline{u}\right)  }{\sigma_{0}}\simeq q_{0}h_{rms}%
\frac{E^{\ast}}{\sigma_{0}}\exp\left(  \frac{-\overline{u}}{\gamma h_{rms}%
}\right)  -\frac{1}{2}\times\left[  Erfc\left(  \frac{\overline{u}-\epsilon
}{\sqrt{2}h_{rms}}\right)  -Erfc\left(  \frac{\overline{u}}{\sqrt{2}h_{rms}%
}\right)  \right]  \label{magic}%
\end{equation}
which obviously results in a pull off finding the minimum as a function of
$\overline{u}$.

Notice that
\begin{equation}
\beta=\frac{E^{\ast}}{\sigma_{0}}=\frac{2E^{\ast}h_{0}^{2}}{\varepsilon
_{0}V^{2}} \label{beta}%
\end{equation}
for low voltage goes to infinity whereas for very high voltage goes to zero,
suggesting the material encounters a rigid wall. Notice that in the standard
case of van der Waals forces, $\frac{E^{\ast}}{\sigma_{0}}\simeq\frac{E^{\ast
}}{\Delta\gamma/\epsilon}=\frac{\epsilon}{l_{a}}$ where $l_{a}=\Delta
\gamma/E^{\ast}$ defines a characteristic adhesion length (for Lennard Jones
with crystals of the same material is $l_{a}\simeq0.05\epsilon$ or
$\frac{\epsilon}{l_{a}}\simeq20$). In that case, $\sigma_{0}=\frac
{\Delta\gamma}{\epsilon}=\frac{l_{a}E^{\ast}}{\epsilon}=0.05E^{\ast}$, namely
the theoretical strength. In the electroadhesion case, $\beta=\frac{2E^{\ast
}h_{0}^{2}}{\varepsilon_{0}V^{2}}$ replaces $\frac{\epsilon}{l_{a}}$ and
depending on the voltage, can span very many orders of magnitude, probably
more than the van der Waals case.

In the generic case, when both van der Waals adhesion and electroadhesion are
at play, one would need to sum their effects.

Fig.2 shows the abrupt decay in pull-off values for constant $\beta=20$,
stickiness is defined (for example) when -$\sigma_{\min}/\sigma_{0}=10^{-8}$
finding this by numerical routines. Moreover, Fig.3 shows the effect of
increasing $\beta$, that is to decrease the voltage, which clearly reduces the
tensile tractions.

\begin{center}
$%
\begin{array}
[c]{c}%
{\includegraphics[
height=3.1194in,
width=5.0548in
]%
{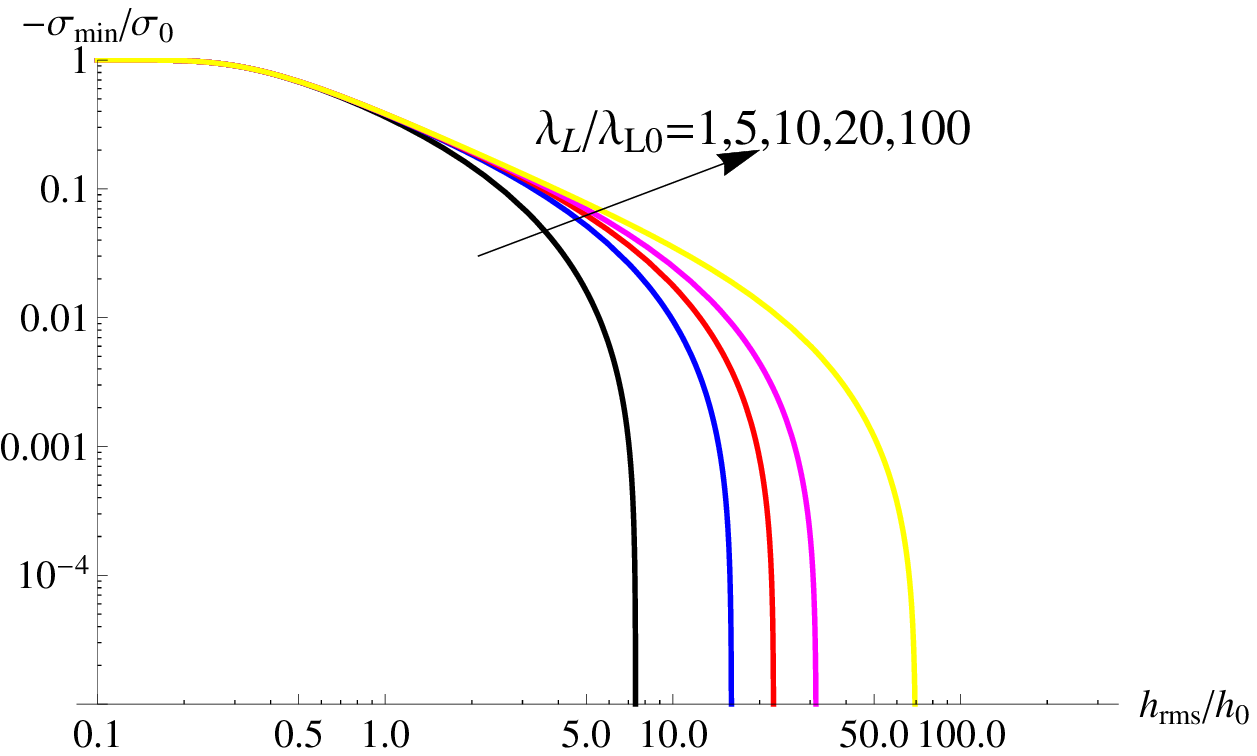}%
}
\\
\end{array}
$

Fig.2- Curves of decay of pull-off normalized pressure -$\sigma_{\min}%
/\sigma_{0}$ as a function of normalized rms roughness amplitude
$h_{rms}/h_{0}$ for constant voltage parameter $\beta=\frac{E^{\ast}}%
{\sigma_{0}}=\frac{2E^{\ast}h_{0}^{2}}{\varepsilon_{0}V^{2}}=20$. Here, the
reference long wavelength cutoff $\lambda_{L0}=\frac{q_{0}}{2\pi}%
=2048\epsilon$ and the curves shift to the right with increasing $\lambda
_{L}/\lambda_{L0}=10^{0},10^{1}...10^{5}$.

$%
\begin{array}
[c]{c}%
{\includegraphics[
height=3.0761in,
width=5.0548in
]%
{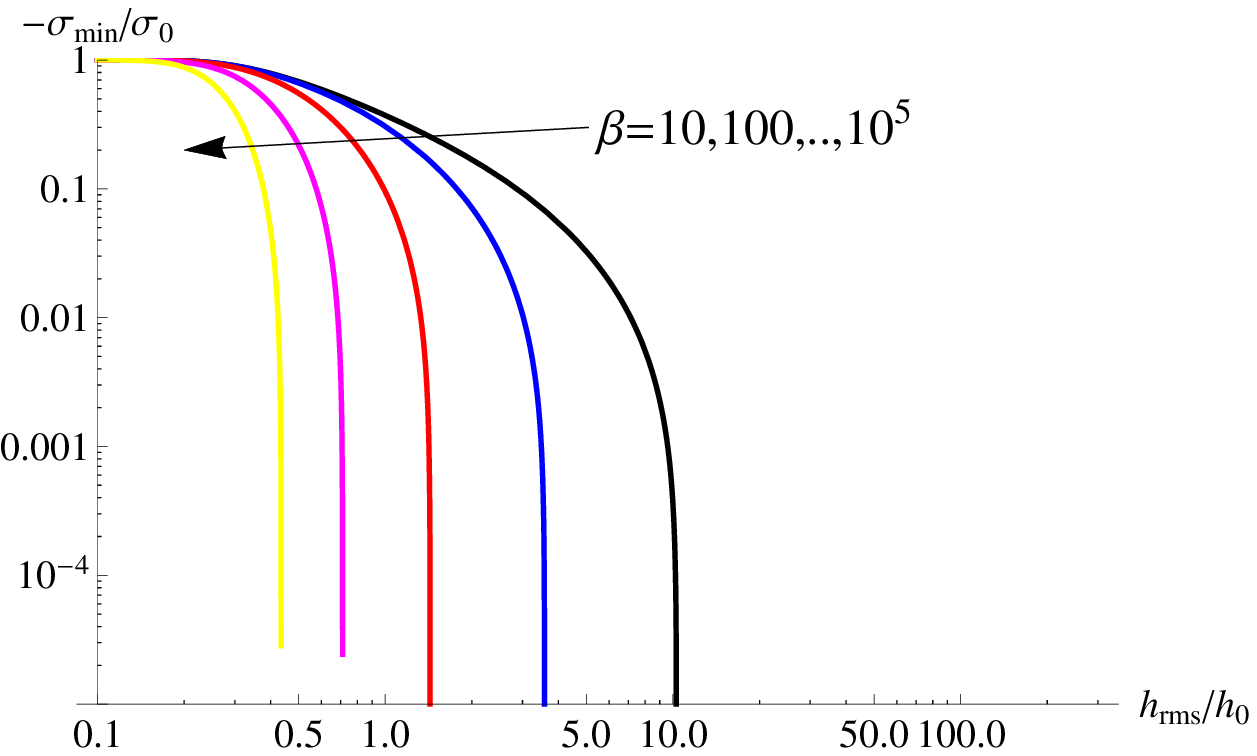}%
}
\\
\end{array}
$

Fig.3- Curves of decay of pull-off normalized pressure -$\sigma_{\min}%
/\sigma_{0}$ as a function of normalized rms roughness amplitude
$h_{rms}/h_{0}$ for constant voltage parameter $\beta=\frac{E^{\ast}}%
{\sigma_{0}}=\frac{2E^{\ast}h_{0}^{2}}{\varepsilon_{0}V^{2}}=10,...,10^{5}$.
Here, the reference long wavelength cutoff $\lambda_{L}=\lambda_{L0}%
=\frac{q_{0}}{2\pi}=2048h_{0}$

\end{center}

\subsection{Area - load}

Persson's theory (2018) has a prediction for the proportion of actual contact
at a given nominal pressure which reads%

\begin{equation}
\frac{A_{rep}}{A_{0}}=\operatorname{erf}(\frac{\sqrt{\pi}}{2}\frac{p_{rep}%
}{\sigma_{rough}}) \label{Persson2}%
\end{equation}
where $\sigma_{rough}=E^{\ast}h_{rms}^{\prime}/2$ where $h_{rms}^{\prime}$ is
the rms slope of the surface, and $p_{rep}$ can be estimate as a function of
$u$ from (\ref{Persson1}). This could be used to estimate the friction load as
proportional to the repulsive contact area.

\begin{center}

\end{center}

\subsection{Stickiness criterion}

The results show that the pull-off traction is principally determined by
$h_{rms},q_{0}$ and upon increasing the "magnification" of the surface,
$\zeta=q_{1}/q_{0}$, converges rapidly, as in the adhesionless load-separation
relation (\ref{Persson1}).

Summarizing, we can obtain similarly to the recent paper by Ciavarella (2019)
for normal van der Waals adhesion, that for stickiness%
\begin{equation}
h_{rms}<\left(  0.6\frac{\sigma_{0}h_{0}}{E^{\ast}}\lambda_{L}\right)
^{0.5}=0.775\sqrt{\frac{h_{0}\lambda_{L}}{\beta}}%
\end{equation}
where we recognize that the dimensionless parameter previously defined $\beta$
is an "electroadhesive" stickiness.

\section{Idea 2 - Energy method}

Alternatively, we could use the energy balance in the Persson-Tosatti's theory
for van der Waals adhesion. Persson \&\ Tosatti (2001) argue with a energy
balance between the state of full contact and that of complete loss of contact
that the effective energy available at pull-off with a rough interface is
$\ $
\begin{equation}
\Delta\gamma_{eff}=\frac{A}{A_{0}}\Delta\gamma-\frac{U_{el}}{A_{0}}
\label{PerssonTosatti}%
\end{equation}
where $A$ is an area in full contact, increased with respect to the nominal
one $A_{0}$, because of an effect of roughness-induced increase of contact
area, $\frac{A}{A_{0}}>1$. Also, $U_{el}$ is the elastic strain energy stored
in the halfspace having roughness with isotropic power spectrum $C\left(
q\right)  $ when this is squeezed flat\footnote{Notice we use the original
Persson's convention and notation for $C\left(  q\right)  $ and not Dalvi et
al. (2019) which is $C^{iso}\left(  q\right)  =4\pi^{2}C\left(  q\right)  $.}
\begin{equation}
\frac{U_{el}\left(  \zeta\right)  }{A_{0}}=\frac{\pi E^{\ast}}{2}\int_{q_{0}%
}^{q_{1}}q^{2}C\left(  q\right)  dq=E^{\ast}l\left(  \zeta\right)  \label{Uel}%
\end{equation}
where we have integrated over wavevectors in the range $q_{0}$, $q_{1}$, and
$E^{\ast}=E/\left(  1-\nu^{2}\right)  $ is the plane strain elastic modulus,
where $\nu$ is Poisson's ratio. We have introduced in (\ref{Uel}) a length
scale $l\left(  \zeta\right)  $ where $\zeta=q_{1}/q_{0}$ is the so called "magnification".

\bigskip

With electroadhesion, we could sum the $\Delta\gamma=\frac{\varepsilon_{0}%
}{2h_{0}}V^{2}$ term, resulting in (let us remove the van der Waals
contribution)%
\begin{equation}
\Delta\gamma_{eff}=\frac{\varepsilon_{0}}{2h_{0}}V^{2}-\frac{U_{el}}{A_{0}}
\label{PT1}%
\end{equation}

\bigskip We can then obtain a new "Persson-Tosatti" stickiness criterion, by
imposing $\Delta\gamma_{eff}=0$ in (\ref{PT1}) obtaining in terms of roughness
amplitude, and using results in Ciavarella (2019), the condition (using
$\beta$ from (\ref{beta})) is obtained%
\begin{equation}
h_{rms}<\sqrt{\frac{h_{0}\lambda_{L}}{\beta}\frac{2H-1}{\pi H}}
\label{PTcriterion}%
\end{equation}
which for $H=0.8$ (the most typical Hurst exponent, see Persson, 2014),
becomes
\begin{equation}
h_{rms}<0.5\sqrt{\frac{h_{0}\lambda_{L}}{\beta}}%
\end{equation}
which compares well with the other criterion obtained with BAM, except the
threshold is 0.5 instead of 0.775.

We can rewrite the criteria \bigskip in the form{ }%
\begin{align}
\frac{h_{rms}}{h_{0}}  &  <0.5\sqrt{\frac{1}{\beta}\frac{\lambda_{L}}{h_{0}}%
}\text{\qquad;\qquad\ Persson-Tosatti}\label{c1}\\
\frac{h_{rms}}{h_{0}}  &  <0.775\sqrt{\frac{1}{\beta}\frac{\lambda_{L}}{h_{0}%
}}\text{\qquad;\qquad\ BAM} \label{c3}%
\end{align}
and a comparison is shown in Fig.4, where Persson-Tosatti is reported in blue
solid line, BAM as black solid line. Clearly, considering the two criteria
have so different origin, it is remarkable that they give so qualitatively
close results. All results are for $H=0.8$ ($D=2.2$), which is the most
typical fractal dimension (Persson, 2014).

\begin{center}
$%
\begin{array}
[c]{cc}%
{\includegraphics[
height=3.1756in,
width=5.0548in
]%
{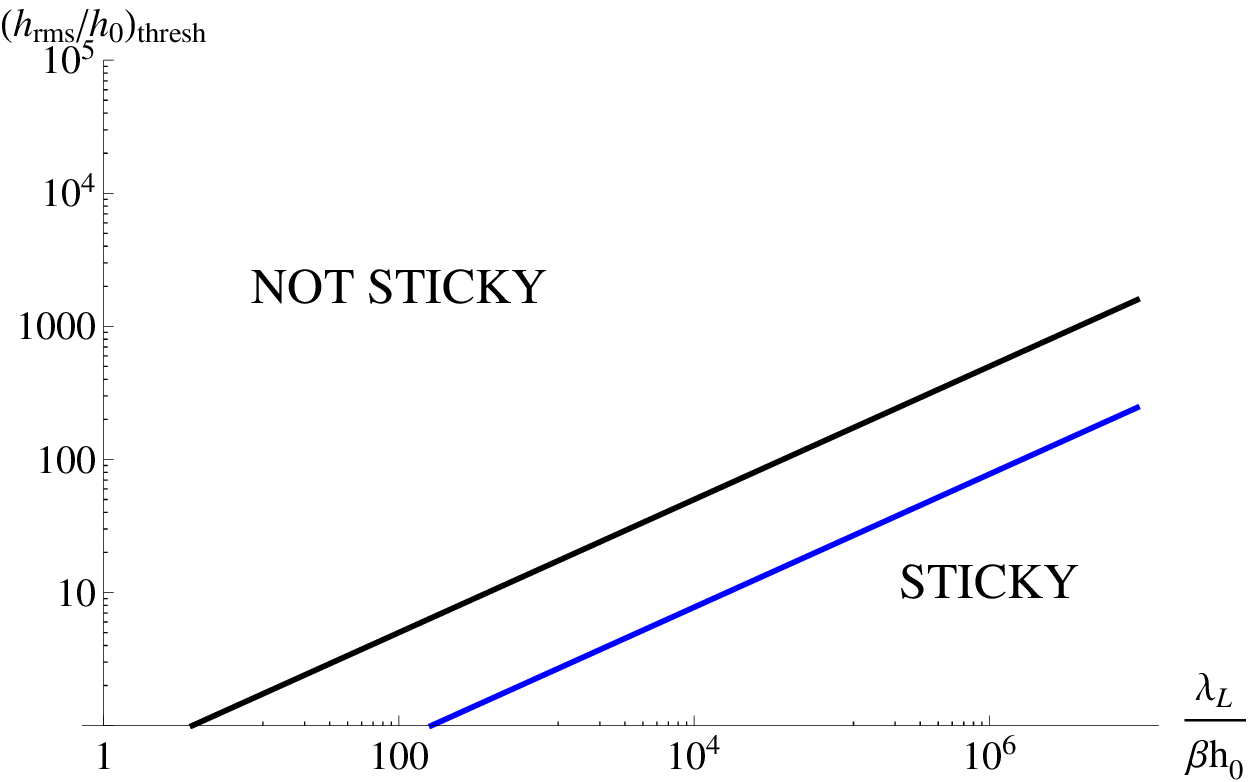}%
}
&
\end{array}
$

Fig.4. A comparison of the three derived stickiness criteria: Persson-Tosatti
(blue line) (\ref{c1}), BAM (black solid line) (\ref{c3}), in terms of the rms
amplitude of roughness. All results are for $H=0.8$ ($D=2.2$), which is the
most typical fractal dimension (Persson, 2014).
\end{center}

\section{Conducting solids in the contact area}

Most solids have a non-zero electric conductivity and we need to take into
account that an electric current will flow through the asperity contact
regions when an electric voltage is applied between the solids: also, the
voltage drop $V$ over the contacting interface will depend on the electric
conductivities 1 and 2 of the solids and on the contact resistance and Persson
(2018, eqt.16) finds approximately%
\begin{equation}
\Delta V=\frac{V}{1+4\frac{d_{0}}{h_{rms}}p_{rep}}%
\end{equation}
where
\begin{equation}
d_{0}=\kappa\left(  \frac{d_{1}}{\kappa_{1}}+\frac{d_{2}}{\kappa_{2}}\right)
\end{equation}
and $\kappa_{i}$ is the electric conductivity of solid $i$ and $\kappa$ is the
effective electric conductivity of the two solids. Hence, when the repulsive
pressure increases, we have much less voltage drop available. Substituting
this result in the previous results, we have a complete picture.

\bigskip

\section{Conclusions}

We have derived a simplified theory for electroadhesion, stemming from Persson
(2018), for power law PSD spectrum, based on BAM, and we have also derived
corresponding a stickiness criterion, which we further compared with an
alternative derivation based on energy balance, stemming from
Persson-Tosatti's (2001), ideas. We have defined a new dimensionless parameter
for electroadhesive stickiness.

\section{References}

Almqvist, C. Campana, N. Prodanov, and B. N. J. Persson, \textquotedblleft
Interfacial separation between elastic solids with randomly rough surfaces:
Comparison between theory and numerical techniques,\textquotedblright\ J.
Mech. Phys. Solids 59, 2355 (2011).

Barber, J. R. \textquotedblleft Bounds on the electrical resistance between
contacting elastic rough bodies,\textquotedblright\ Proc. R. Soc. A 459, 53 (2003).

Ciavarella, M., Joe, J., Papangelo, A., Barber, JR. (2019) The role of
adhesion in contact mechanics. J. R. Soc. Interface, 16, 20180738

Ciavarella, M. (2018) A very simple estimate of adhesion of hard solids with
rough surfaces based on a bearing area model. Meccanica, 1-10. DOI 10.1007/s11012-017-0701-6

Ciavarella, M. (2019). Universal features in" stickiness" criteria for soft
adhesion with rough surfaces. arXiv preprint arXiv:1908.06380.

\bigskip Dalvi, S., Gujrati, A., Khanal, S. R., Pastewka, L., Dhinojwala, A.,
\& Jacobs, T. D. (2019). Linking energy loss in soft adhesion to surface
roughness. arXiv preprint arXiv:1907.12491.

\bigskip

Persson B. N. J. (2018), The dependency of adhesion and friction on
electrostatic attraction, The Journal of Chemical Physics 148, 144701 (2018);
doi: 10.1063/1.5024038

Popov, V. L., \& Hess, M. (2018). Voltage induced friction in a contact of a
finger and a touchscreen with a thin dielectric coating. arXiv preprint arXiv:1805.08714.

Persson, B. N. J. (2007). Relation between interfacial separation and load: a
general theory of contact mechanics. Physical review letters, 99(12), 125502.

Persson, B. N. J., \& Tosatti, E. (2001). The effect of surface roughness on
the adhesion of elastic solids. The Journal of Chemical Physics, 115(12), 5597-5610

\bigskip Persson, B. N. J. (2014). On the fractal dimension of rough surfaces.
Tribology Letters, 54(1), 99-106.

Vardar, Y., Guclu, B. and Basdogan, C. \textquotedblleft Effect ofwaveform on
tactile perception by electrovibration displayed on touch
screens,\textquotedblright\ IEEE Trans. Haptics 10, 488 (2017).

\end{document}